\documentclass[12pt]{article}
\usepackage[left=1in,right=1in,top=1in,bottom=1in]{geometry}
\usepackage{graphicx}
\usepackage{float}
\usepackage{cite}
\usepackage{amsfonts}
\usepackage{amssymb}
\usepackage{amsmath}
\usepackage{relsize}
\usepackage{color}

\def\gtwid{\mathrel{\raise.3ex\hbox{$>$\kern-.75em\lower1ex\hbox{$\sim$}}}}
\def\ltwid{\mathrel{\raise.3ex\hbox{$<$\kern-.75em\lower1ex\hbox{$\sim$}}}}
\def\square{\kern1pt\vbox{\hrule height 1.2pt\hbox{\vrule width 1.2pt\hskip 3pt
   \vbox{\vskip 6pt}\hskip 3pt\vrule width 0.6pt}\hrule height 0.6pt}\kern1pt}
\newcommand{\p}{\partial}
\newcommand{\meanvac}[1]{\left\langle \Omega\left| #1 \right| \Omega \right\rangle}

\newcommand{\dd}{\mathcal{D}}
\newcommand{\red}[1]{{\color{red} #1 \color{black}}}
\newcommand{\blue}[1]{{\color{blue} #1 \color{black}}}

\begin{document}
\begin{titlepage}

\begin{flushright}
UFIFT-QG-23-07
\end{flushright}

\vskip 4cm

\begin{center}
{\bf Large Inflationary Logarithms in a Nontrivial Nonlinear Sigma Model}
\end{center}

\vskip 1cm

\begin{center}
C. Litos$^{\dagger}$, R. P. Woodard$^{*}$ and B. Yesilyurt$^{\ddagger}$
\end{center}

\begin{center}
\it{Department of Physics, University of Florida, \\
Gainesville, FL 32611, UNITED STATES}
\end{center}

\vskip 1cm

\begin{center}
ABSTRACT
\end{center}

Loops of inflationary gravitons are known to induce large temporal and
spatial logarithms which can cause perturbation theory to break down.
Nonlinear sigma models possess the same kind of derivative interactions
and induce the same sorts of large logarithms, without the complicated
index structure and potential gauge problem. Previous studies have
examined models with zero field space curvature which can be reduced to
free field theories by local, invertible field redefinitions. Here we
study a model which cannot be so reduced and still shows the same sorts
of large logarithms. We compute the evolution of the background at 1-loop
and 2-loop orders, and we find the 1-loop $\beta$ and $\gamma$ functions. 

\begin{flushleft}
\end{flushleft}

\vspace{3cm}

\begin{flushleft}
$^{\dagger}$ e-mail: c.litos@ufl.edu \\
$^{*}$ e-mail: woodard@phys.ufl.edu \\
$^{\ddagger}$ e-mail: b.yesilyurt@ufl.edu 
\end{flushleft}

\end{titlepage}
\section{Introduction}

The background geometry of cosmology is characterized by scale factor $a(t)$,
Hubble parameter $H(t)$, and first slow roll parameter $\epsilon(t)$,
\begin{equation}
ds^2 = -dt^2 + a^2(t) d\vec{x} \!\cdot\! d\vec{x} \qquad \Longrightarrow \qquad
H(t) \equiv \frac{\dot{a}}{a} \quad , \quad \epsilon(t) \equiv -\frac{\dot{H}}{H^2}
\; . \label{geometry}
\end{equation}
The accelerated expansion of inflation ($H(t) > 0$ with $0 \leq \epsilon(t) < 1$)
rips virtual scalars and gravitons out of the vacuum. This is the basis for the
primordial spectra of scalars \cite{Mukhanov:1981xt} and tensors 
\cite{Starobinsky:1979ty}. At some level these quanta must interact among
themselves and with other particles. Gravitons are especially interesting because
their couplings are universal and because their tensor structure allows them to 
mediate effects which scalars cannot. For example, on de Sitter background 
($\epsilon(t) = 0$) one loop gravitons modify the plane wave mode function 
$u(t,k)$ of gravitational radiation \cite{Tan:2021lza} and the gravitational 
response $\Psi(t,r)$ to a point mass \cite{Tan:2022xpn} to,
\begin{eqnarray}
u(t,k) &\!\!\! = \!\!\!& u_0(t,k) \Biggl\{1 + \frac{16 G H^2}{3\pi} \ln^2(a)
+ O(G^2)\Biggr\} \; , \qquad \label{ugrav} \\
\Psi(t,r) &\!\!\! = \!\!\!& -\frac{G M}{a r} \Biggl\{1 + \frac{103 G}{15 \pi a^2 r^2}
- \frac{8 G H^2}{\pi} \ln^3(a) + O(G^2)\Biggr\} \; . \qquad \label{Psigrav}
\end{eqnarray}
Similar results have been obtained for the corrections of inflationary gravitons
to fermions \cite{Miao:2006gj}, to electrodynamics \cite{Glavan:2013jca,
Wang:2014tza}, and to massless, minimally coupled scalars \cite{Glavan:2021adm}.

A fascinating aspect of these corrections is that the steady production of 
inflationary gravitons endows them with a secular growth which must eventually 
overwhelm the loop-counting parameter $G H^2$ provided inflation persists long
enough. One must develop a nonperturbative resummation technique in order to 
evolve past that point. Nonlinear sigma models provide a useful testing ground
for developing such a technique because they possess the same sorts of derivative
interactions as gravitons, and induce the same sorts of large logarithms, without
the complicating tensor structures or the potential for gauge dependence 
\cite{Tsamis:2005hd,Kitamoto:2010et,Kitamoto:2011yx,Kitamoto:2018dek}. A
successful technique has been devised through combining a variant of 
Starobinsky's stochastic formalism \cite{Starobinsky:1986fx,Starobinsky:1994bd} 
with a variant of the renormalization group \cite{Miao:2021gic,Woodard:2023rqo}.
Even better, the technique can be generalized from de Sitter to an arbitrary
cosmological background (\ref{geometry}) which has undergone primordial
inflation \cite{Kasdagli:2023nzj}. Applying this technique shows that the
large scales of primordial inflation can be transmitted to late time
\cite{Woodard:2023cqi}.

The obvious next step is generalizing the resummation technique from nonlinear
sigma models to quantum gravity. This seems to be entirely possible, and works
for the one graviton loop correction on which it has been checked 
\cite{Glavan:2021adm}. However, our purpose here is to clear up a worry 
concerning the sorts of nonlinear sigma models on which the resummation 
technique has so far been applied. Specifically, both of those models can be 
reduced to free theories by means of local invertible field redefinitions,
which means that their flat space S-matrices are unity by Borchers Theorem 
\cite{Borchers:1960}. That in no way precludes interesting evolutions for 
the scalar backgrounds, and for the 1-particle states, and it was these
evolutions which suggested and confirmed the resummation technique. One might
still worry that the technique only works for models which can be reduced to
free theories. The purpose of this paper is to demonstrate that the resummation
technique applies for a model whose flat space $S$-matrix is nontrivial.

This paper contains five sections, of which this Introduction is the first.
In Section 2 we describe the old and new models, and we give the Feynman 
rules. Section 3 computes the time-evolving background at 1-loop and 2-loop
orders. It also computes and renormalizes the 1PI (one-particle-irreducible)
2-point and 3-point functions at 1-loop order. The resummation technique is
applied in Section 4. Our conclusions comprise Section 5.

\section{The Model}

The purpose of this section is to define the model and give those of its 
Feynman rules which are required for our work. We begin by presenting the
old models and explaining why their $S$-matrices are trivial. Then a new
model is proposed by making a small variation which preserves the lowest
order interactions.

The resummation technique was developed based on work with two nonlinear
sigma models \cite{Miao:2021gic}. One of these was based on a single field.
All models of this sort can be made free by a local, invertible field 
redefinition,
\begin{equation}
\mathcal{L} = -\frac12 f^2(\Phi) \partial_{\mu} \Phi \partial_{\nu} \Phi
g^{\mu\nu} \sqrt{-g} \quad , \quad d\Psi \equiv f(\Phi) d\Phi \quad
\Longrightarrow \quad \mathcal{L} = -\frac12 \partial_{\mu} \Psi \partial_{\nu}
\Psi g^{\mu\nu} \sqrt{-g} \; . \label{1field}
\end{equation}
A second model was based on two fields,
\begin{equation}
\mathcal{L}_{\rm old} = -\frac12 \partial_{\mu} A \partial_{\nu} A g^{\mu\nu} 
\sqrt{-g} - \frac12 \Bigl(1 + \frac{\lambda}{2} A\Bigr)^2 \partial_{\mu} B 
\partial_{\nu} B g^{\mu\nu} \sqrt{-g} \; . \label{oldmodel}
\end{equation}
Although it was not initially realized, this model can be also reduced to a 
theory of two free scalars by making a local, invertible field 
redefinition,\footnote{We thank Arkady Tseytlin for this observation.}
\begin{eqnarray}
X &\!\!\! \equiv \!\!\!& \frac{2}{\lambda} \Bigl(1 + \frac{\lambda}{2} A\Bigr)
\cos\Bigl( \frac{\lambda}{2} B\Bigr) \; , \\
Y &\!\!\! \equiv \!\!\!& \frac{2}{\lambda} \Bigl(1 + \frac{\lambda}{2} A\Bigr)
\sin\Bigl( \frac{\lambda}{2} B\Bigr) \; .
\end{eqnarray}
Hence the flat space S-matrices of (\ref{1field}) and (\ref{oldmodel}) are both
unity by Borchers Theorem \cite{Borchers:1960}. That in no way precludes these 
models from manifesting interesting evolutions of their backgrounds and of the
single-particle kinematics. 

\begin{figure}[ht]
    \centering
    \includegraphics[scale= 0.65]{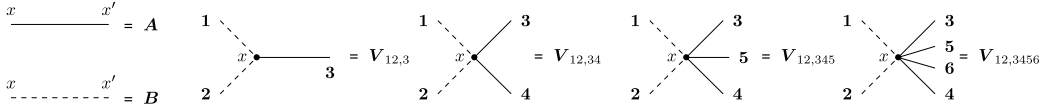}
    \caption{\footnotesize Primitive interactions of the bare Lagrangian 
    (\ref{newmodel}). A lines are solid and $B$ lines are dashed.}
    \label{fig:FRules}
\end{figure}

To be certain that the resummation technique does not rely on having a trivial
S-matrix we devised a slight modification of the 2-field model (\ref{oldmodel})
whose field space curvature implies that it cannot be reduced to a fee theory,
\begin{equation}
\mathcal{L}_{\rm new} = -\frac12 \partial_{\mu} A \partial_{\nu} A g^{\mu\nu} 
\sqrt{-g} - \frac12 \Bigl(1 + \frac{\lambda}{4} A\Bigr)^4 \partial_{\mu} B 
\partial_{\nu} B g^{\mu\nu} \sqrt{-g} \; . \label{newmodel}
\end{equation} 
The 3-point coupling in this model is identical to that of the old model
(\ref{oldmodel}), and the 4-point coupling has the same field content with
the old numerical coefficient of $\frac18$ replaced by $\frac{3}{16}$. There
are additional 5-point and 6-point interactions which make only simple 
contributions to the diagrams we evaluate in section 3, 
\begin{equation}
\frac{1}{2} \Bigl(1 \!+\! \frac{\lambda}{4} A \Bigr)^4 (\p B)^2 - \frac{1}{2} 
(\p B)^2 = \frac{\lambda}{2} A (\p B)^2 + \frac{3 \lambda^2}{16} A^2 (\p B)^2 
+ \frac{\lambda^3}{32} A^3 (\p B)^2 + \frac{\lambda^4}{512} A^4 (\p B)^2 . 
\label{feynexpansion}
\end{equation}
A diagrammatic representation of the Feynman rules is shown in 
Figure~\ref{fig:FRules}.

\begin{figure}[ht]
    \centering
    \includegraphics[scale= 0.85]{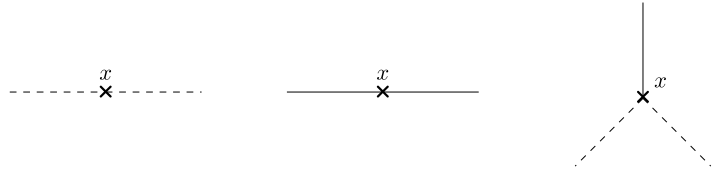}
    \caption{\footnotesize Diagrammatic representation of the 1-loop counterterms
    we require. The first two diagrams renormalize the $A$ and $B$ self-masses
    while the rightmost diagram renormalizes the vertex function. Recall that $A$
    lines are solid and $B$ lines are dashed.}
    \label{fig:allcounter}
\end{figure}

The new model (\ref{newmodel}) is no more renormalizable than the old one 
(\ref{oldmodel}). Hence divergences must be subtracted, order-by-order, using
BPHZ counterterms (Bogoliubov-Parasiuk-Hepp-Zimmermann \cite{Bogoliubov:1957gp,
Hepp:1966eg,Zimmermann:1968mu,Zimmermann:1969jj}). The 1-loop counterterms we
require are shown in Figure~\ref{fig:allcounter}.
\noindent
The first and second diagrams renormalize the $A$ and $B$ self-masses and 
correspond to the counterterm,
\begin{eqnarray}
\lefteqn{\Delta \mathcal{L}^{(2)} = -\frac12 C_{1A^2} \square A \square A 
\sqrt{- g} - \frac12 C_{2A^2} R \, \partial_{\mu} A \partial_{\nu} A g^{\mu\nu} 
\sqrt{-g} } \nonumber \\
& & \hspace{6cm} -\frac12 C_{1B^2} \square B \square B \sqrt{-g} - 
\frac12 C_{2B^2} R \, \partial_{\mu} B \partial_{\nu} B g^{\mu\nu} \sqrt{-g} 
\; . \qquad \label{AandB2cterms}
\end{eqnarray}
The 3rd diagram is required to renormalize the 3-point vertex and corresponds
to,
\begin{eqnarray}
\lefteqn{\delta \mathcal{L}^{(3)} = - \frac{1}{2} C_{1AB^2} \Box A \p_\mu B 
\p_\nu B g^{\mu\nu} \sqrt{-g} - C_{2AB^2} \p_\mu A \p_\nu B \Box B g^{\mu\nu} 
\sqrt{-g} } \nonumber \\
& & \hspace{5cm} - \frac{1}{2} C_{3AB^2} A \Box B \Box B \sqrt{-g} - 
\frac{1}{2} C_{4A B^2} R A \p_\mu B \p_\nu B  g^{\mu\nu} \sqrt{-g} \ . \qquad
\label{vertcount}
\end{eqnarray}
Here $R = D (D-1) H^2$ is the Ricci scalar. Section 3 will determine the various 
coefficients in expressions (\ref{AandB2cterms}-\ref{vertcount}) as functions of 
$\lambda$, $D$ and the dimensional regularization scale $\mu$.

In $D$ spacetime dimensions, the propagators of both fields obey the equation,
\begin{equation}
\p^\mu \left[ a^{D-2} \p_\mu i \Delta(x;x') \right] \equiv 
\mathcal{D} i \Delta(x;x') = i \delta^D (x-x'). \label{propeq}
\end{equation}
The solution consists of a de Sitter invariant part plus a de Sitter breaking 
logarithm \cite{Onemli:2002hr,Onemli:2004mb},
\begin{equation}
i\Delta(x;x') = F\Bigl(\mathcal{Y}(x;x')\Bigr) + k \ln(aa') \qquad , \qquad 
k \equiv \frac{H^{D-2}}{(4\pi)^{D/2}} \frac{\Gamma(D \!-\! 1)}{\Gamma(\frac{D}2)}. \label{propsol}
\end{equation}
Here $\mathcal{Y}(x;x') \equiv a a' H^2 \Delta x^2(x;x') \equiv a a' H^2 (x - x')^2$ 
is the de-Sitter length function, and the first derivative of $F(\mathcal{Y}(x;x'))$ 
obeys, 
\begin{eqnarray}
\lefteqn{F'(\mathcal{Y}) = - \frac{H^{D-2}}{4 (4\pi)^{D/2}} \Biggl\{ 
\Gamma\Bigl(\frac{D}{2}\Bigr) \left( \frac{4}{\mathcal{Y}} \right)^{\frac{D}2} + 
\Gamma\Bigl(\frac{D}{2} \!+\! 1\Bigr) \left( \frac{4}{\mathcal{Y}}\right)^{\frac{D}2 - 1} }
\nonumber \\
& & \hspace{4.5cm} + \sum_{n=0}^\infty \left[ \frac{\Gamma(n \!+\! \frac{D}{2} \!+\! 2)}{
\Gamma(n \!+\! 3)} \left( \frac{\mathcal{Y}}{4} \right)^{n - \frac{D}{2} + 2} - 
\frac{\Gamma(n \!+\! D)}{\Gamma(n \!+\! \frac{D}{2} \!+\!1)} 
\left( \frac{\mathcal{Y}}{4} \right)^{n} \right] \Biggr\} . \qquad \label{Fprime}
\end{eqnarray} 
In dimensional regularization the coincidence limits of the propagator and its
derivatives are,
\begin{eqnarray}
i \Delta(x;x) = k \left[ - \pi \cot\left( \frac{D \pi}{2} \right) + 2 \ln (a) \right] 
\qquad & , & \qquad \p_\mu i \Delta(x;x')\Bigl\vert_{x=x'} = k H a \delta^0_{\ \mu} 
\; , \qquad \label{coinc1} \\
\p_\mu \p'_\nu i \Delta(x;x')\Bigl\vert_{x'=x} = - \left( \frac{D-1}{D} \right) 
k H^2 g_{\mu\nu} \qquad & , & \qquad \p_\mu i \Delta(x;x) = 2 k H a \delta^0_{\ \mu} 
\; . \label{coinc2}
\end{eqnarray}

We close by commenting on notation. Because the de Sitter metric $g_{\mu\nu} = 
a^2 \eta_{\mu\nu}$ is conformal to the Minkowski metric $\eta_{\mu\nu}$, we adopt
a notation where $\partial_{\mu}$ stands for $\frac{\partial}{\partial x^{\mu}}$,
no matter what sort of tensor it acts upon. Further, we raise and lower its indices
using the Minkowski metric, $\partial^{\mu} \equiv \eta^{\mu\nu} \partial_{\nu}$.
And we define $\partial^2 \equiv \eta^{\mu\nu} \partial_{\mu} \partial_{\nu}$. To 
save space we sometimes write coordinate arguments of the metric and its scale factor
using a subscript or a superscript, as in $\sqrt{-g(x)} \equiv \sqrt{-g_x} \equiv 
a_x^{D}$ and $g^{\mu\nu}(y) \equiv g^{\mu\nu}_{y} \equiv a^{-2}_{y} \eta^{\mu\nu}$.
The same notation applies to derivatives, as in $\frac{\partial}{\partial z^{\mu}}
\equiv \partial^{z}_{\mu}$.

\section{Explicit Results}

The purpose of this section is to carry out the same explicit calculations
for (\ref{newmodel}) that were done for the old model (\ref{oldmodel})
\cite{Miao:2021gic,Woodard:2023rqo}. We begin with the 1-loop and 2-loop
expectation values of $A(x)$. Next, the self-masses of $A$ and $B$ are
computed and renormalized at 1-loop order. Finally, we evaluate the 1-loop
vertex function.

\subsection{The 1-Loop and 2-Loop Background}

\begin{figure}[ht]
    \centering
    \includegraphics[scale= 0.65]{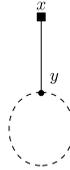}
    \caption{\footnotesize 1-loop contribution to the expectation value of $A(x)$.}
    \label{fig:vevA1loop}
\end{figure}

We start with the 1-loop expectation value of $A(x)$ whose diagrammatic representation 
is shown in Figure~\ref{fig:vevA1loop}. 
\noindent Because the 3-point couplings of the old and new models agree, this diagram 
is unchanged from the old result \cite{Miao:2021gic}, 
\begin{equation}
\boxed{\meanvac{A(x)} = \frac{\lambda H^2}{16\pi^2} \left[ \ln(a) - \frac{1}{3} + 
\frac{1}{3a^3} \right] + \mathcal{O}(\lambda^3)}. \label{1loopA}
\end{equation}

\begin{figure}[ht]
    \centering
    \includegraphics[scale= 0.65]{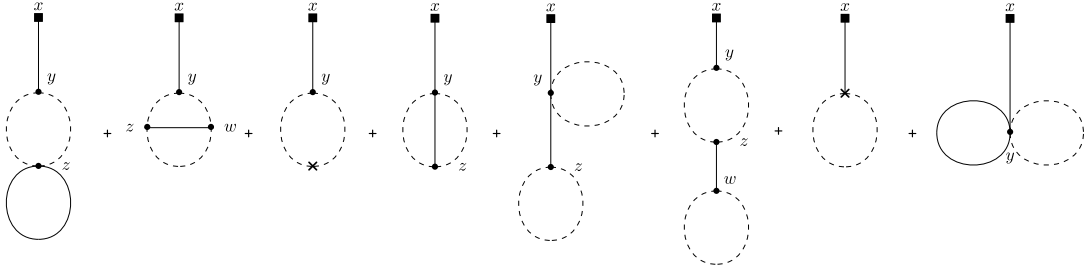}
    \caption{\footnotesize Contributions to the expectation value of $A(x)$ at 2-loop order.}
    \label{fig:VEVAall}
\end{figure}

The 2-loop contributions are shown in Figure~\ref{fig:VEVAall}. 
The first 7 diagrams (which we label $A_{2a}$ through $A_{2g}$) were calculated in 
\cite{Miao:2021gic}. We need only include a relative factor of $\frac32$ for each 4-point 
coupling to express their final contributions, at leading logarithm order, as multiples of 
the factor $S \equiv \lambda^3 H^4 \ln^2(a)/2^{10}\pi^4$,
\begin{eqnarray}
A_{2a} \longrightarrow -3 \cdot S \quad , \quad A_{2b} \longrightarrow +8 \cdot S 
\quad & , & \quad A_{2c} \longrightarrow 0 \cdot S \quad , \quad A_{2d} \longrightarrow 
-6 \cdot S \; , \qquad \\
A_{2e} \longrightarrow +\frac{3}{2} \cdot S \quad , \quad A_{2f} \longrightarrow -2 
\cdot S \quad & , & \quad  A_{2g} \longrightarrow 0 \cdot S \; . \qquad 
\end{eqnarray}

Only the last diagram, $A_{2h}$ is really new. It has a symmetry factor of $\frac{1}{4}$
and its initial expression is,
\begin{equation}
A_{2h} = \frac{-i}{4} \cdot \frac{3\lambda^3}{8} \cdot \int d^D y \ \sqrt{-g(y)} 
\, g^{\mu\nu}(y) \cdot i \Delta(x;y) \cdot i\Delta(y;y) \cdot \p_{\mu}^{z} \p_{\nu}^y 
i \Delta(y;z)\Bigl\vert_{y=z} \; .
\end{equation}
The two coincidence limits can be read from (\ref{coinc1}-\ref{coinc2}) to give,
\begin{equation}
A_{2h} =\frac{3i \lambda^3}{2^5} (D-1) k H^2 I_2,
\end{equation}
where $I_2$, and its leading logarithm result, was given in \cite{Woodard:2023cqi},
\begin{equation}
I_2 \equiv \int d^Dy \sqrt{-g(y)} i \Delta(x;y) i \Delta(y;y) \longrightarrow
- \frac{i}{24\pi^2} \times \ln^2(a).
\end{equation}
This is finite and does not require renormalization, so we can set $D=4$,
\begin{equation}
A_{2h} \longrightarrow \frac{3}{2} \cdot S.
\end{equation}
Adding this to the leading logarithm results for the first seven diagrams gives,
\begin{eqnarray}
A_{2a} \longrightarrow -3 \cdot S \quad , \quad A_{2b} \longrightarrow +8 \cdot S 
\quad & , & \quad A_{2c} \longrightarrow 0 \cdot S \quad , \quad A_{2d} \longrightarrow 
-6 \cdot S \; , \qquad \label{indcont1} \\ 
A_{2e} \longrightarrow +\frac{3}{2} \cdot S \quad , \quad A_{2f} \longrightarrow -2 
\cdot S \quad & , & \quad  A_{2g} \longrightarrow 0 \cdot S \quad , \quad A_{2h} 
\longrightarrow +\frac{3}{2} \cdot S \; . \qquad \label{indcont2}
\end{eqnarray}
At leading logarithm order the eight diagrams of Figure~\ref{fig:VEVAall} sum to zero,
so our result for the expectation value of $A$ is,
\begin{equation}
\boxed{\meanvac{A(x)} = \frac{\lambda H^2 \ln(a)}{16\pi^2} \Bigl[ 1 + 0 \Bigr] + 
\mathcal{O}(\lambda^4)} \; . \label{2loopA}
\end{equation}

\subsection{The 1-Loop Self-Masses}

We now move on to the 1-loop self-masses. Absorbing the divergences of these reveals 
curvature-dependent field strength renormalizations $Z_{A}$ and $Z_{B}$ from the
terms proportional to $C_{2A^2}$ and $C_{2B^2}$ in expression (\ref{AandB2cterms}).
These give the $\gamma$ functions that will be used in the Renormalization Group (RG) 
analysis of Section 4.

\begin{figure}[ht]
    \centering
    \includegraphics[scale= 0.65]{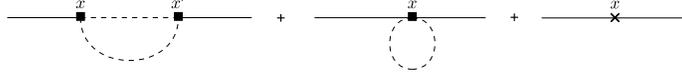}
    \caption{\footnotesize 1-loop contributions to the $A$ self mass $-i M_A^2(x;x')$.}
    \label{fig:M2A}
\end{figure}

The 1-loop contributions to $-i M^2_{A}(x;x')$ are shown in Figure~\ref{fig:M2A}. 
The first diagram is unchanged from the old model, and the second diagram is just 
$\frac32$ times the previous result \cite{Miao:2006gj},
\begin{eqnarray}
-i M_{A_3}^2(x;x') & \!\!\! = \!\!\! & \frac{(-i\lambda)^2}{2} \Biggl\{ \frac{1}{4} 
\dd \dd' \Bigl[ i \Delta(x;x') \Bigr]^2 \nonumber \\ 
& & \hspace{3cm} - \frac{1}{2} \dd \Bigl[ i \Delta(x;x)i \delta^D(x \!-\! x') \Bigr] 
- k H a^{D-1} \p_0 i \delta^D(x \!-\! x') \Biggr\} , \qquad \label{A3} \\
-i M^2_{A_4}(x;x') & \!\!\! = \!\!\! & - \frac{i3\lambda^2}{8} \delta^D(x \!-\! x') 
\cdot -(D \!-\! 1) k H^2 a^D \; . \label{A4}
\end{eqnarray}
The counterterm is $i$ times the second variation of (\ref{AandB2cterms}), 
\begin{equation}
-i M^2_{Ac}(x;x') = -C_{1 A^2} \dd \dd' \left[ \frac{i\delta^D(x \!-\! x')}{(aa')^{\frac{D}2}} 
\right] + C_{2 A^2} \p^\mu \Bigl[ R a^{D-2} \p_\mu i \delta^D(x \!-\! x') \Bigr] . \label{Ac}
\end{equation}
Because only (\ref{A3}) diverges in dimensional regularization, and this diagram is
identical to the old model, so too are the coefficients $C_{A_1}$ and $C_{A_2}$,
\begin{equation}
C_{1 A^2} = - \frac{\lambda^2 \mu^{D-4}}{32 \pi^{\frac{D}2}} 
\frac{\Gamma(\frac{D}{2} \!-\! 1)}{2 (D \!-\! 3) (D \!-\! 4)} \quad , \quad 
C_{2 A^2} = \frac{\lambda^2 \mu^{D-4}}{4(4\pi)^{\frac{D}2}} \frac{\Gamma (D \!-\! 1)}{
\Gamma(\frac{D}2)} \frac{\pi \cot(\frac{D\pi}{2})}{D (D \!-\! 1)} . \label{C2A2}
\end{equation}

\begin{figure}[ht]
    \centering
    \includegraphics[scale= 0.65]{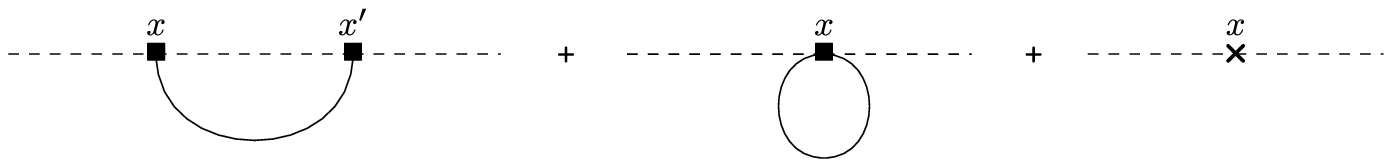}
    \caption{\footnotesize 1-loop contributions to the $B$ self-mass $-i M_B^2(x;x')$.}
    \label{fig:M2B}
\end{figure}

We move to the $B$ self-mass, whose 1-loop contributions are shown in 
Figure~\ref{fig:M2B}.
As before, the 3-point diagram is unchanged from the old model and the 4-point diagram
is just the result from the old model times $\frac32$,
\begin{equation}
-i M^2_{B_3} = \frac{(i\lambda)^2}{2} \dd \dd' \Bigl[ i \Delta(x;x') \Bigr]^2 - 
(i\lambda)^2 \p^\mu \p'^\rho \Bigl[ (aa')^{D-2} \p_\mu i \Delta(x;x') \p'_\rho i 
\Delta(x;x') \Bigr] \label{B3}
\end{equation}
\begin{equation}
-i M^2_{B_4} = - \frac{3 \lambda^2 k \pi \cot(\frac{D \pi}2) \dd}{8} 
\left[ i \delta^D(x \!-\! x') \right] + \frac{3\lambda^2 H^2 \p^\mu}{32\pi^2}
\Bigl[ \ln(a) a^2 \p_\mu i \delta^4(x \!-\! x') \Bigr] + \mathcal{O}(D \!-\! 4) . 
\label{B4}
\end{equation}
The counterterm is $i$ times the second variation of (\ref{AandB2cterms}) with 
respect to $B$,
\begin{equation}
- i M^2_{B_c}(x;x') = - C_{1B^2} \dd \dd' \left[ \frac{i\delta^D(x \!-\! x')}{(aa')^{\frac{D}2}} 
\right] + C_{2B^2} \p^\mu \Bigl[ R a^{D-2} \p_\mu i \delta^D(x \!-\! x') \Bigr] . \label{Bc}
\end{equation}
Multiplying the divergences from (\ref{B4}) by $\frac32$ and combining with those
from (\ref{B3}) gives,
\begin{equation}
C_{1B^2} = - \frac{\lambda^2 \mu^{D-4}}{16\pi^{\frac{D}2}} \frac{\Gamma(\frac{D}{2} \!-\! 1)}{
2 (D \!-\! 3) (D \!-\! 4)} ,
\end{equation}
\begin{equation}
C_{2B^2} = \frac{3\lambda^2 \mu^{D-4}}{8(4\pi)^{\frac{D}2}} \frac{\Gamma(D \!-\! 1)}{
\Gamma(\frac{D}{2})} \frac{\pi \cot(\frac{D\pi}{2})}{D (D \!-\! 1)} - 
\frac{\lambda^2 \mu^{D-4}}{32\pi^{\frac{D}2}} \frac{\Gamma(\frac{D}{2} \!-\! 1)}{2 
(D \!-\! 3) (D \!-\! 4)} \frac{D \!-\! 2}{D \!-\! 1} . \label{C2B2}
\end{equation}

\subsection{The 1-Loop Vertex Function}

In this subsection we first isolate the primitive divergences of the 3-point vertex
$-i V(x;y;z)$ at 1-loop order. These divergences are removed by the counterterm 
(\ref{vertcount}), which determines the coefficients $C_{1A B^2}$, $C_{2A B^2}$,
$C_{3A B^2}$ and $C_{4A B^2}$. We regard the $C_{4A B^2}$ counterterm as a 
curvature-dependent renormalization of the bare 3-point coupling and infer the 
corresponding beta function as,
\begin{equation}
\delta \lambda = C_{4 AB^2} \!\times\! R + O(\lambda^5) \qquad \Longrightarrow
\qquad \beta \equiv \frac{\partial \delta \lambda}{\partial \ln(\mu)} \; .
\label{beta}
\end{equation}

\begin{figure}[ht]
    \centering
    \includegraphics[scale= 0.65]{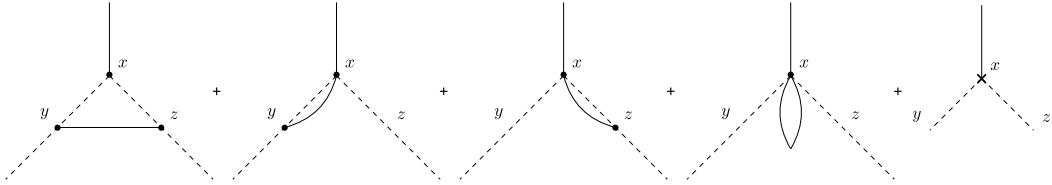}
    \caption{\footnotesize 1-loop contributions to the $ABB$ vertex $-i V(x;y;z)$.}
    \label{fig:Vertex}
\end{figure}

The tree order vertex can be inferred from the Feynman rules, 
\begin{equation}
-iV_0(x;y;z) = -i\lambda\sqrt{-g(x)} \, g^{\mu\nu}_x \p_\mu \delta^D(x-y) \p_\nu 
\delta^D(x \!-\! z). \label{V0}
\end{equation}
The various 1-loop contributions are shown in Figure~\ref{fig:Vertex}.
Because the leftmost diagram involves only 3-point couplings, it is unchanged from
the old model \cite{Woodard:2023rqo},
\begin{eqnarray}
\lefteqn{-i V_{1 a}(x ; y ; z)=\frac{i \lambda^3}{2} \mathcal{D}_x \partial_\sigma^y 
\partial_\beta^z \Biggl\{\sqrt{-g_y} \, g_y^{\rho \sigma} \partial_\rho^y i 
\Delta(x ; y) \sqrt{-g_z} \, g_z^{\alpha \beta} \partial_\alpha^z i \Delta(x ; z) 
i \Delta(y ; z) \Biggr\} } \nonumber \\
& & \hspace{5.5cm} + \frac{\lambda^3}{4} \mathcal{D}_y \mathcal{D}_z \Biggl\{
\Bigl[i \Delta(y ; z)\Bigr]^2 \Bigl[\delta^D(x \!-\! y) + \delta^D(x \!-\! z)
\Bigr] \Biggr\} \nonumber \\
& & \hspace{1.2cm} - \frac{\lambda^3}{2} \partial_\sigma^y \partial_\beta^z \Biggl\{
\sqrt{-g_y} \, g_y^{\rho \sigma} \partial_\rho^y i \Delta(y ; z) \sqrt{-g_z} \, 
g_z^{\alpha \beta} \partial_\alpha^z i \Delta(y ; z) \Bigl[\delta^D(x \!-\! y)
+ \delta^D(x \!-\! z) \Bigr] \Biggr\} . \qquad \label{firstcontr}
\end{eqnarray}
The 2nd and 3rd diagrams involve a single 4-point coupling and are therefore $\frac32$
times the results of the old model \cite{Woodard:2023rqo},
\begin{eqnarray}
\lefteqn{-i V_{1b}(x;y;z) = \frac{3\lambda^3}{8} \mathcal{D}_y \Biggl\{\sqrt{-g_x} \,
g_x^{\mu \nu} \partial_\mu^x \Bigl[i \Delta(x ; y)\Bigr]^2 \partial_\nu 
\delta^D(x \!-\! z) \Biggr\} } \nonumber \\
& & \hspace{3cm} + \frac{3\lambda^3}{4} \partial_\nu^z \partial_\beta^y \Biggl\{
\sqrt{-g_z} \, g_z^{\mu \nu} \partial_\mu^z i \Delta(z ; y) \sqrt{-g_y} \, 
g_y^{\alpha \beta} \partial_\alpha^y i \Delta(z ; y) \delta^D(x \!-\! z) \Biggr\} ,
\qquad \label{secondcontr} \\
\lefteqn{-i V_{1 c}(x;y;z) = \frac{3\lambda^3}{8} \mathcal{D}_z \Biggl\{\sqrt{-g_x} \,
g_x^{\mu \nu} \partial_\mu^x[i \Delta(x ; z)]^2 \partial_\nu \delta^D(x \!-\! y)
\Biggr\} } \nonumber \\
& & \hspace{3cm} + \frac{3\lambda^3}{4} \partial_\sigma^y \partial_\beta^z \Biggl\{
\sqrt{-g_y} \, g_y^{\rho \sigma} \partial_\rho^y i \Delta(y ; z) \sqrt{-g_z} \,
g_z^{\alpha \beta} \partial_\alpha^z i \Delta(y ; z) \delta^D(x \!-\! y) \Biggr\} . 
\qquad \label{thirdcontr}
\end{eqnarray} 

Before considering the 4th diagram, we combine and reduce expressions 
(\ref{firstcontr}-\ref{thirdcontr}). Because the last two differ from 
the old model, the reduction is more complicated. Adding all three terms
and performing some judicious partial integrations gives, 
\begin{eqnarray}
\lefteqn{-i V_{1abc} = \frac{i \lambda^3}{2} \mathcal{D}_x \partial^\rho_y 
\partial^\alpha_z \Bigl\{(a_y a_z)^{D-2} i \Delta(y ; z)\partial_\rho^y i \Delta(x ; y) 
\partial_\alpha^z i \Delta(x ; z) \Bigr\} }\nonumber \\
& & \hspace{0cm} + \frac{3\lambda^3}{8} \mathcal{D}_y \p_z^\alpha \Bigl\{ a_z^{D-2} 
\Bigl[i \Delta(y;z) \Bigr]^2 \p_\alpha^z \delta^D(x \!-\! z) \Bigr\} + \frac{3\lambda^3}{8} 
\mathcal{D}_z \p_y^\alpha \Bigl\{ a_y^{D-2} \Bigl[i \Delta(y;z) \Bigr]^2 \p_\alpha^y 
\delta^D(x \!-\! y) \Bigr\} \nonumber \\
& & \hspace{4cm} - \frac{\lambda^3}{8} \mathcal{D}_y \mathcal{D}_z \Bigl\{ \Bigl[
i\Delta(y;z) \Bigr]^2  \Bigl[\delta^D(x \!-\! y) + \delta^D(x \!-\!z) \Bigr] \Bigr\} 
\nonumber \\ 
& & \hspace{3cm} + \frac{\lambda^3}{4} \partial^{\rho}_y \partial^{\alpha}_z \Bigl\{
(a_y a_z)^{D-2} \partial_\rho^y i \Delta(y ; z) \partial_\alpha^z i \Delta(y ; z) 
\Bigl[\delta^D(x \!-\! y) + \delta^D(x \!-\! z)\Bigr] \Bigr\} . \qquad \label{v1prim}
\end{eqnarray}
After considerable manipulations explained in the Appendix A expression the divergent 
part of (\ref{v1prim}) can be brought to the form,
\begin{equation}
-iV_{1abc}(x;y;z) \longrightarrow -\frac{i\lambda^3 \Gamma(\frac{D}{2} \!+\! 1) 
\mu^{D-4} H^2 a_x^2}{32 \pi^{\frac{D}2} (D \!-\! 3) (D \!-\! 4)} \, \p^\mu 
\delta^D(x \!-\! y) \p_\mu \delta^D(x \!-\! z) -i \widetilde{V}(x;y;z) , \label{V1abc}
\end{equation}
where $-i \widetilde{V}(x;y;z)$ consists of higher derivative divergences, 
\begin{eqnarray}
\lefteqn{-i \widetilde{V}(x;y;z) = \frac{i\lambda^3 \mu^{D-4}}{16\pi^{\frac{D}2}} 
\frac{\Gamma(\frac{D}{2} \!-\! 1)}{D (D \!-\! 3) (D \!-\! 4)} \dd_x 
\left[ \frac{\p_\mu \delta^D(x \!-\! y) \p^\mu \delta^D(x \!-\! z)}{a_x^{D-2}} 
\right] } \nonumber \\
& & \hspace{-0.5cm} + \frac{5i \lambda^3 \mu^{D-4} \Gamma(\frac{D}{2} \!-\! 1)}{
128 \pi^{\frac{D}2} (D \!-\! 3) (D \!-\!4)} \Biggl\{ \dd_y \p_x^\mu \left[ 
\frac{\delta^D(x \!-\! y) \p_\mu \delta^D(x \!-\! z)}{a_x^{D-2}} \right] 
+ \dd_z \p_x^\mu \left[ \frac{\p_\mu \delta^D(x \!-\! y) \!\cdot\! 
\delta^D(x \!-\! z)}{a_x^{D-2}} \right] \Biggr\} \nonumber \\ 
& & \hspace{5.5cm} - \frac{\lambda^3 \Gamma(\frac{D}{2} \!-\! 1) \mu^{D-4}}{64
\pi^{\frac{D}2} (D \!-\! 3) (D \!-\! 4)} \dd_y \dd_z \left[ \frac{i 
\delta^D(x \!-\! z) \delta^D(x \!-\! y)}{a_x^{2D-4}} \right] . \qquad 
\end{eqnarray}

We call the new diagram $-i V_{1d}(x;y;z)$ and its contribution is,
\begin{equation}
-i V_{1d} = -i  \frac{3\lambda^3}{16} \sqrt{-g(x)} \, g^{\mu\nu}(x) \p_\mu 
\delta^D(x \!-\! y) \p_\nu \delta^D(x \!-\! z)  \cdot i \Delta(x;x) \; .
\end{equation}
This is just the bare vertex (\ref{V0}) times the simple factor 
$\frac{3 \lambda^2}{16} i\Delta(x;x)$. Recall that the coincident propagator is 
given in equation (\ref{coinc1}),
\begin{eqnarray}
-iV_{1d} &\!\!\! = \!\!\!& -i \frac{3 \lambda^3}{16} a_x^{D-2} \, \p^\mu \delta^D(x \!-\! y) 
\p_\mu \delta^D(x \!-\! z) k \left[ - \pi \cot\Bigl(\frac{D\pi}{2}\Bigr) + 2\ln(a)\right]
\; , \qquad \\
&\!\!\! = \!\!\!& -i \frac{3 \lambda^3 k}{16} \!\times\! -\pi \cot\Bigl( \frac{D \pi}{2}
\Bigr) \!\times\! a_x^2 \, \partial^{\mu} \delta^D(x \!-\! y) \partial_{\mu} \delta^D(x \!-\! z)
+ O(D \!-\! 4) \; . \qquad
\end{eqnarray}
Adding the primitive divergences from (\ref{V1abc}) to this gives,
\begin{eqnarray}
\lefteqn{-i V_{1{\rm pdiv}}(x;y;z) =-i \frac{\lambda^3 H^2 a_x^2}{8} \, \p^\mu 
\delta^D(x \!-\! y) \p_\mu \delta^D(x \!-\! z) } \nonumber \\
& & \hspace{4cm} \times \left[\frac{3 k}{2 H^2} \!\times\! -\pi \cot\Bigl(
\frac{D\pi}{2}\Bigr) + \frac{\mu^{D-4} \Gamma(\frac{D}{2} \!+\! 1)}{4 \pi^{\frac{D}2}
(D \!-\! 3) (D \!-\! 4)} \right] - i\Tilde{V}(x;y;z) . \qquad \label{V1prim} 
\end{eqnarray}

The primitive divergences (\ref{V1prim}) are absorbed by the third variation of 
(\ref{vertcount}),
\begin{eqnarray}
\lefteqn{\frac{i \delta^3 S_{AB^2}[A,B]}{\delta A(x) \delta B(y) \delta B(z)} = -i C_{1AB^2} 
\dd_x \left[ \frac{\p^\mu \delta^D(x \!-\! y) \p_\mu \delta^D(x \!-\! z)}{a_x^2} \right] }
\nonumber \\
& & \hspace{0cm} + i C_{2AB^2} \dd_y \p^\mu_x \left[ \frac{\delta^D(x \!-\! y) \p_\mu 
\delta^D(x \!-\! z)}{a_x^2} \right] + i C_{2AB^2} \dd_z \p^\mu_x \left[ \frac{\p_\mu 
\delta^D(x \!-\! y) \!\times\! \delta^D(x \!-\! z)}{a_x^2} \right] \nonumber \\ 
& & \hspace{1.7cm} -i C_{3AB^2} \dd_y \dd_z \left[ \frac{\delta^D(x \!-\! y) 
\delta^D(x \!-\! z)}{a_x^D} \right] -i C_{4AB^2} R a_x^{2} \, \p^\mu \delta^D(x \!-\! y) 
\p_\mu \delta^D(x \!-\! z) \; . \qquad \label{V1count}
\end{eqnarray}
Adding (\ref{V1count}) to (\ref{V1prim}) and demanding regularity as $D \to 4$ 
implies,
\begin{eqnarray}
C_{1AB^2} &\!\!\! = \!\!\!& \frac{\lambda^3 \mu^{D-4} \, \Gamma(\frac{D}{2} \!-\! 1)}{
16 \pi^{\frac{D}2} D (D \!-\! 3) (D \!-\! 4)} \; , \qquad \\
C_{2AB^2} &\!\!\! = \!\!\!& -\frac{5 \lambda^3 \mu^{D-4} \, \Gamma(\frac{D}{2} \!-\! 1)}{
128 \pi^{\frac{D}2} (D \!-\! 3) (D \!-\! 4)} \; , \qquad \\ 
C_{3AB^2} &\!\!\! = \!\!\!& -\frac{\lambda^3 \mu^{D-4} \, \Gamma(\frac{D}{2} \!-\! 1)}{
64 \pi^{\frac{D}2} (D \!-\! 3)(D \!-\! 4)} \; , \qquad \\
C_{4AB^2} &\!\!\! = \!\!\!& \frac{\lambda^3 \mu^{D-4}}{32 \pi^{\frac{D}2} D (D \!-\! 1)} 
\left[ \frac{3\pi \cot(\frac{D\pi}{2}) \Gamma(D \!-\! 1)}{8 \Gamma(\frac{D}{2})} 
- \frac{\Gamma(\frac{D}2 \!+\! 1)}{(D \!-\! 3) (D \!-\! 4)} \right] \; . \qquad \label{C4}
\end{eqnarray}
Because we have suppressed the finite contributions the renormalized result will not 
be given but let us take note of the fact that logarithms of $\mu$ come in the form
$\ln(\frac{\mu a}{H})$. We can regard $C_{4A B^2} \times R = \delta \lambda$ as a 
curvature-dependent renormalization. The associated 1-loop beta function (\ref{beta}) is,
\begin{equation}
\boxed{\beta = \frac{\p \delta \lambda}{\p \ln(\mu)} = -\frac{\lambda^3 H^2}{64\pi^2}
+ O(\lambda^5) } \; . \label{betanew}
\end{equation}

\section{Summing the Logarithms}

Here we demonstrate that the results of the previous section for the new model
(\ref{newmodel}) can be explained using the same combination of Starobinsky's 
stochastic formalism and the renormalization group that worked for the old model
(\ref{oldmodel}). We begin by deriving the curvature-dependent effective potential
and showing that it explains the evolution of the background. We next apply our
explicit results for the 1-loop counterterms to work out the curvature-dependent
renormalization group corrections.

\subsection{Curvature-Dependent Effective Potential}

The shift symmetry of the field $B(x)$ evident in (\ref{newmodel}) prevents it 
from developing an effective potential. However, $A(x)$ does acquire one from 
integrating out the differentiated $B$ fields from the $A$ field equation in
the presence of a constant $A(x) = A_0$ background,
\begin{equation}
\frac{\delta S[A,B]}{\delta A(x)} = \partial_{\mu} \Bigl[\sqrt{-g} \, g^{\mu\nu} 
\partial_{\nu} A \Bigr] - \frac{\lambda}{2} \Bigl(1 \!+\! \frac{\lambda}{4} A \Bigr)^3 
\p_\mu B \p_\nu B  g^{\mu\nu} \sqrt{-g} = 0 \; . \label{b3}
\end{equation}
One can see from the Lagrangian (\ref{newmodel}) that a constant $A(x) = A_0$
background just renormalizes the $B$ field strength,
\begin{equation}
\Bigl\langle \Omega \Bigl\vert T\Bigl[ B(x) B(x') \Bigr] \Bigr\vert \Omega 
\Bigr\rangle_{A = A_0} = \frac{i \Delta(x;x')}{(1 + \frac{\lambda}{4} A_0)^4} \; .
\end{equation}
Hence the $A$ equation can be recognized as that of a scalar potential model,
\begin{eqnarray}
\lefteqn{\frac{\delta S[A,B]}{\delta A(x)} \longrightarrow \partial_{\mu} \Bigl[\sqrt{-g} 
\, g^{\mu\nu} \partial_{\nu} A \Bigr] - \frac{\lambda}{2} \Bigl(1 \!+\! \frac{\lambda}{4} 
A \Bigr)^3 \times \frac{\sqrt{-g} \, g^{\mu\nu} \partial_{\mu} \partial'_{\nu}
i \Delta(x;x') \vert_{x'=x}}{(1 \!+\! \frac{\lambda}{4} A)^4} \; , } \label{force} \\
& & \hspace{0cm} = \partial_{\mu} \Bigl[\sqrt{-g} \, g^{\mu\nu} \partial_{\nu} A \Bigr]
+ \frac{(\frac{D-1}{2}) \lambda k H^2 \sqrt{-g}}{1 \!+\! \frac{\lambda}{4} A} \equiv 
\partial_{\mu} \Bigl[\sqrt{-g} \, g^{\mu\nu} \partial_{\nu} A \Bigr] -  
V'_{\rm eff}(A) \sqrt{-g} \; . \qquad \label{Vdef}
\end{eqnarray}
Note that we have employed expression (\ref{coinc2}) to evaluate the coincidence limit
of the doubly differentiated propagator. This is free of divergences, so we can set 
$D=4$ to find, 
\begin{equation}
V'_{\text{eff}}(A) = - \frac{3\lambda H^4}{16\pi^2} \Bigl(1 + \frac{\lambda}{4} A 
\Bigr)^{-1} \qquad \Longrightarrow \qquad V_{\rm eff}(A) = - \frac{3 H^4}{4 \pi^2} 
\ln\Bigl(1 \!+\! \frac{\lambda}{4} A\Bigr) \; . \label{effectivepotential}
\end{equation}
We see that the new model's curvature-dependent effective potential 
(\ref{effectivepotential}) is almost the same as the old model's result of 
$-\frac{3 H^4}{8 \pi^2} \ln(1 + \frac{\lambda}{2} A)$ \cite{Miao:2021gic}. 

Starobinsky long ago showed how to sum the leading logarithms of a scalar potential
model like (\ref{Vdef}) \cite{Starobinsky:1986fx,Starobinsky:1994bd}. The leading 
logarithms contained in correlators of the quantum field $A(t,\vec{x})$ turn out to 
agree, to all orders \cite{Tsamis:2005hd}, with those of stochastic random field 
$\mathcal{A}(t,\vec{x})$ which obeys the Langevin equation,
\begin{equation}
3 H \Bigl[ \dot{\mathcal{A}} - \dot{\mathcal{A}}_0\Bigr] = V'_{\rm eff}(\mathcal{A})
\; . \label{Langevin}
\end{equation}
The ``stochastic jitter'' in this equation is supplied by the time derivative of the 
infrared-truncated, free field expansion of $A(t,\vec{x})$,
\begin{equation}
\mathcal{A}_0 (t,\vec{x}) \equiv \int_{H}^{aH} \!\!\!\! \frac{d^3k}{(2 \pi)^3} 
\frac{H}{\sqrt{2 k^3}} \Bigl\{ \alpha(\vec{k}) e^{i \vec{k} \cdot \vec{x}} + 
\alpha^{\dagger}(\vec{k}) e^{-i \vec{k}\cdot \vec{x}} \Bigr\} \quad , \quad 
\Bigl[\alpha(\vec{k}) , \alpha^{\dagger}(\vec{k}') \Bigr]
= (2\pi)^3 \delta^3(\vec{k} \!-\! \vec{k}') \; . \label{jitter} 
\end{equation}
If we turn off the stochastic jitter then equation (\ref{Langevin}) is simple to
solve, adopting the initial condition $\mathcal{A}(0,\vec{x}) = 0$ and noting that
$\ln(a) = H t$,
\begin{equation}
\mathcal{A}(t,\vec{x}) \Bigl\vert_{\mathcal{A}_0 = 0} = \frac{4}{\lambda} \left[ 
\sqrt{1 \!+\! \frac{\lambda^2 H^2 \ln(a)}{32\pi^2}} - 1 \right] = 
\frac{\lambda H^2 \ln(a)}{16 \pi^2} - \frac{\lambda^3 H^4 \ln^2(a)}{2^{11} \pi^4} 
+ O(\lambda^5) \; . \label{Asol}
\end{equation}
Because it is easier to fluctuate down the potential than up, we expect that the
effect of adding stochastic jitter is to accelerate the evolution of $\mathcal{A}$
down its potential. Solving (\ref{Langevin}) perturbatively gives,
\begin{equation}
\mathcal{A} = \mathcal{A}_0 + \frac{\lambda H^2 \ln(a)}{16 \pi^2} - 
\frac{\lambda^2 H^3}{2^6 \pi^2} \!\! \int_0^t \!\!\! dt' \mathcal{A}_0(t',\vec{x}) -
\frac{\lambda^3 H^4 \ln^2(a)}{2^{11} \pi^4} + \frac{\lambda^3 H^3}{2^8 \pi^2} \!\!
\int_0^t \!\!\! dt' \mathcal{A}^2_0(t',\vec{x}) + O(\lambda^4) \; . \label{withjitter}
\end{equation}
By taking the expectation value of the previous equation, and using the mode sum
(\ref{jitter}) to conclude,
\begin{equation}
\Bigl\langle \Omega \Bigl\vert \mathcal{A}_0(t,\vec{x}) \Bigr\vert \Omega \Bigr\rangle 
= 0 \qquad , \qquad \Bigl\langle \Omega \Bigl\vert \mathcal{A}_0^2(t,\vec{x}) \Bigr\vert
\Omega \Bigr\rangle = \frac{H^2 \ln(a)}{4 \pi^2} \; ,
\end{equation}
we find, 
\begin{equation}
\Bigl\langle \Omega \Bigl\vert \mathcal{A}(t,\vec{x}) \Bigr\vert \Omega \Bigr\rangle = 
\frac{\lambda H^2 \ln(a)}{16 \pi^2} + O(\lambda^5) \; .
\label{stochasticresult}
\end{equation}
This is exactly the result (\ref{2loopA}) we got by explicit computation, which is
impressive confirmation of the stochastic prediction.

\subsection{Curvature-Dependent Renormalization Group}

In a theory with derivative interactions not all of the large logarithms derive from
stochastic effects. Some of them arise instead from the incomplete cancellation between
primitive divergences --- which have no $D$-dependent powers of the scale factor ---
and counterterms --- which inherit a $a^D$ from the measure factor $\sqrt{-g}$,
\begin{equation}
\Bigl( \frac{H^{D-4}}{D \!-\! 4}\Bigr) - \Bigl(\frac{\mu^{D-4} a^{D-4}}{D \!-\! 4}\Bigr)
= -\ln\Bigl( \frac{ \mu a}{H} \Bigr) + O(D \!-\! 4) \; . \label{RGsource}
\end{equation}
These logarithms can be recovered using the renormalization group, which follows factors
of $\ln(\mu)$. Because we are only interested in cases where the factors of $\ln(a)$ are 
not suppressed by inverse powers of the scale factor, the counterterms of concern are 
those which can be viewed as curvature-dependent renormalizations of bare parameters 
\cite{Miao:2021gic}. Of the four counterterms (\ref{vertcount}) required to renormalize 
the vertex function at 1-loop order we have already seen that the contribution
proportional to $C_{4A B^2}$ can be viewed as a curvature-dependent renormalization of
the bare vertex, and we used this to compute the associated beta function (\ref{beta}).
Of the four counterterms (\ref{AandB2cterms}) required to renormalize the 1-loop 
self-masses it is the contributions proportional to $C_{2 A^2}$ and $C_{2 B^2}$ which 
can be regarded as curvature-dependent field strength renormalizations,
\begin{equation}
Z_A \equiv 1 + C_{2 A^2} \!\times\! R + O(\lambda^4) \qquad , \qquad
Z_B \equiv 1 + C_{2 B^2} \!\times\! R + O(\lambda^4) \; . \label{ZAB}
\end{equation}
Our explicit results (\ref{C2A2}) and (\ref{C2B2}) give the associated gamma functions,
\begin{equation}
\boxed{\gamma_{A} \equiv \frac{\p \ln(Z_{A})}{\p \ln(\mu^2)} = \frac{\lambda^2 H^2}{32\pi^2} 
+ O(\lambda^4) \qquad , \qquad \gamma_B \equiv \frac{\p \ln(Z_{B})}{\p \ln(\mu^2)}
= -\frac{\lambda^2 H^2}{64 \pi^2} + O(\lambda^4) \; . } \label{gammas}
\end{equation}

We are ready to investigate the Callan-Symanzik equations for the n-point Green's 
functions $G_n(x_1;x_2;\ldots;x_n;\lambda;\mu)$ of the field $A$,\footnote{1PI 
$n$-point functions obey a similar equation with the last term replaced by $- n 
\gamma_A$.} 
\begin{equation}
\left[ \mu \frac{\p}{\p \mu} + \beta \frac{\p}{\p \lambda} + n\gamma_A \right] 
G_n\Bigl(x_1;x_2;\ldots;x_n;\lambda;\mu\Bigr) = 0 \; . \label{CSeqn}
\end{equation}
This equation can be solved using the method of characteristics. We first find a 
running coupling constant $\bar{\lambda}(\mu)$ which obeys the differential equation
and initial condition,
\begin{equation}
\mu \frac{d\bar{\lambda}}{d\mu}  = - \beta\Bigl(\bar{\lambda}(\mu)\Bigr) \quad , 
\quad \bar{\lambda}(\mu_0) = \lambda \qquad \Longrightarrow \qquad \beta(\lambda) 
\frac{\p \bar{\lambda}}{\p \lambda} = \beta(\bar{\lambda}) \; . \label{newbeta}
\end{equation}
We can then write the solution as,
\begin{equation}
G_n\Bigl(x_1;x_2;\ldots;x_n;\lambda;\mu\Bigr) =  G_n\Bigl(x_1;x_2;\ldots;x_n;
\bar{\lambda}(\mu);\mu_0\Bigr) \!\times\! \exp\left[ -n \!\! \int_{\mu_0}^\mu \!\!
\frac{d\mu'}{\mu'} \gamma\Bigl(\bar{\lambda}(\mu')\Bigr)\right] \; . \label{gensol} 
\end{equation}
Inserting the $\beta$ function (\ref{betanew}) into (\ref{newbeta}), and ignoring
higher loop corrections yields,
\begin{equation}
\bar{\lambda}(\mu) = \frac{\lambda}{\sqrt{1 \!-\! \frac{\lambda^2 H^2}{32 \pi^2} 
\ln(\frac{\mu}{\mu_0})}} \; . \label{lambdabar} 
\end{equation}
Substituting (\ref{lambdabar}) and (\ref{gammas}) into (\ref{gensol}), and again
ignoring higher loop corrections, gives,
\begin{equation}
G_n\Bigl(x_1;x_2;\ldots x_n;\lambda;\mu\Bigr) = G_n\Bigl(x_1;x_2;\ldots;x_n;
\bar{\lambda}(\mu);\mu_0\Bigr) \!\times\! \Bigl[1 \!-\! \frac{\lambda^2 H^2}{32\pi^2} 
\ln\Bigl(\frac{\mu}{\mu_0}\Bigr) \Bigr]^{n} \; . \label{finalsol}
\end{equation}
Similar results could be derived for Green's functions which also, or even exclusively,
involve the field $B(x)$.

Having a negative beta function traditionally means that the theory runs towards
weak coupling in the ultraviolet because logarithms of the scale $\mu$ are associated
with inverse factors of some characteristic momentum in the process. In cosmology we
are interested in how things behave at {\it late times}, and we note from expression
(\ref{RGsource}) that the scale $\mu$ is associated with the scale factor $a(t)$ in
the form $\ln[\mu a/H]$. It should therefore be that having a negative beta function
means the theory evolves towards strong coupling at late times.
 
\section{Conclusion}

In this work, we examined a nontrivial nonlinear sigma model (defined in section 2)
whose loop corrections on de Sitter background show the same large logarithms as 
have been reported from inflationary gravitons \cite{Tan:2021lza,Tan:2022xpn,
Miao:2006gj,Glavan:2013jca,Wang:2014tza,Glavan:2021adm}. Previous work 
\cite{Miao:2021gic,Woodard:2023rqo} on nonlinear sigma models which can be reduced 
to free theories has shown that these large logarithms can be resummed by combining 
a variant of Starobinsky's stochastic formalism \cite{Starobinsky:1986fx,
Starobinsky:1994bd} --- based on curvature-dependent effective potentials --- with 
a variant of the renormalization group --- based on the subset of counterterms 
which can be viewed as curvature-dependent renormalizations of bare parameters. 
Our analysis confirms that these techniques continue to apply for models whose 
S-matrix is nontrivial. In section 3.1, we explicitly computed the evolution of 
the background (\ref{2loopA}) at 1-loop and 2-loop orders. In section 4.1 we 
showed that the resummation techniques correctly predict these results. 

One significant difference associated with a nontrivial S-matrix is that the beta
function does not vanish. The 1-loop beta function (\ref{betanew}) was derived in
section 3.3. The 1-loop gamma functions (\ref{gammas}) were derived in section 3.2,
and combined with the beta function in section 4.2, to solve the Callan-Symanzik 
equation for $n$-point Green's functions (\ref{finalsol}). Because the beta function
(\ref{betanew}) is negative, the running coupling constant $\bar{\lambda}(\mu)$ grows 
with the scale $\mu$ as shown in equation (\ref{lambdabar}). From (\ref{RGsource})
we see that logarithms of $\mu$ are associated with the scale factor $a(t)$ in the 
form $\ln[\mu a(t)/H]$. This implies that the model becomes strongly coupled at late 
times.

An interesting higher loop phenomenon is that there can be mixing of large stochastic
logarithms with large logarithms from the renormalization group. We speculate that
the correct way to include these is to use the renormalization group to improve the
curvature-dependent effective potential and then use that, without any extra RG
corrections. From (\ref{gammas}) we observe that, because $\gamma_A \sim \lambda^2$, 
an RG-improvement to the effective potential (\ref{effectivepotential}) will include 
lowest order corrections of the form $\lambda^3 A$. Those would not engender any
changes in the background evolution at leading logarithm order.

We should comment on how this work can be extended to slow-roll inflation models,
in particular to ultra-slow-roll inflation in which a nearly flat region of the 
potential causes the scalar to almost stop rolling \cite{Tsamis:2003px}. One must
be clear that neither of the nonlinear sigma model scalars $A(x)$ and $B(x)$ 
serves as the inflaton; they are just spectators to inflation driven by some other 
scalar field.\footnote{To make $A$ or $B$ the inflaton would require the introduction
of a potential for them, which is a very substantial modification of the model.} 
In this case exact calculations are no longer possible because we lack the propagators
for a general inflationary background. However, the great thing about the resummation
technique described in section 4 is that it can be implemented on a general 
inflationary background. The stochastic formalism described in section 4.1 relies 
on a curvature-dependent effective potential whose effective force (\ref{force})
derives from the coincidence limit of the doubly differentiated free scalar 
propagator $\partial_{\mu} \partial'_{\nu} i \Delta(x;x')$ in the appropriate 
background. Although we do not have exact expressions for this quantity, very 
good analytic approximations have been developed \cite{Kasdagli:2023nzj}, and one 
can use them to solve the resulting Langevin equation (\ref{Langevin}) numerically 
for any cosmological background which has experienced primordial inflation 
\cite{Woodard:2023cqi}. If the analytic approximations should happen to become
unreliable for a particular expansion history they can easily be improved 
\cite{Brooker:2017kjd}. The situation for the curvature-dependent renormalization
group described in section 4.2 is even better because the coefficients of the
counterterms are universal, independent of the background geometry. 

The infrared modes which engender the large logarithms which we have studied
are fascinating, but experience with gravity \cite{Losic:2004hw,Losic:2005vg,
Losic:2006ht,Unruh:2008zza,Losic:2008ht} shows that one must employ a 
resummation technique which goes beyond linearized order in order to understand
what happens when perturbation theory breaks down. Although our 1-loop and 2-loop
results are obviously perturbative, we emphasize that they were made merely to
check the validity of the resummation technique of section 4, which is fully
nonperturbative. One can see that the technique includes nonlinear effects from
the form (\ref{effectivepotential}) of the curvature-dependent effective potential. 
Note also that its field dependence of $\ln(1 + \tfrac{\lambda}{4} A)$ applies to 
{\it any} cosmological background; only the multiplicative coefficient of 
$-\tfrac{3 H^4}{4 \pi^2}$ changes when the geometry is no longer de Sitter. The 
solution (\ref{Asol}) which results from ignoring stochastic jitter is similarly
nonperturbative --- and it should be generally true that jitter merely serves to
accelerate the rate at which the $A$ background rolls down the effective potential.
Finally, the Renormalization Group formalism sums logarithms which result from 
renormalization (\ref{RGsource}) to all orders in perturbation theory.

\vskip 1cm

\centerline{\bf Acknowledgements}

This work was partially supported by NSF grant PHY-2207514 and by DOE grant 
DE-SC0022148 and by the Institute for Fundamental Theory at the University of Florida.

\appendix 

\section{Reduction of the 1-loop vertex function}\label{betafunctionappendix}

The aim of this appendix is to reduce the last line of equation (\ref{v1prim}) for
$-i V_{1abc}(x;y;z)$,
\begin{equation}
I(x;y;z) \equiv \frac{\lambda^3}{4} \partial^{\rho}_y \partial^{\alpha}_z \Bigl\{
(a_y a_z)^{D-2} \partial_{\rho}^y i \Delta(y;z) \partial_{\alpha}^z i \Delta(y;z) 
\Bigl[\delta^D(x \!-\! y) + \delta^D(x \!-\! z)\Bigr] \Bigr\} \; . \label{Idef}
\end{equation}
Note that $\partial_{\rho}^y i \Delta(y;z) \partial_{\alpha}^z i \Delta(y;z)$ is 
quadratically divergent, so extracting just the divergent part of $I(x;y;z)$ requires 
only the two terms given on the first line of expansion (\ref{Fprime}) for each 
propagator, 
\begin{eqnarray}
\p^y_{\rho} i \Delta(y;z) &\!\!\! = \!\!\!& -\frac{\Gamma(\frac{D}{2})}{2 
\pi^{\frac{D}2} (a_y a_z)^{\frac{D}2 -1}} \left[ \frac{(y \!-\! z)_{\rho}}{
(y \!-\! z)^D} + \frac{a_y H \delta^0_{~\rho}}{2 (y \!-\! z)^{D-2}} + 
\frac{D a_y a_z H^2 (y \!-\! z)_{\rho}}{8 (y \!-\! z)^{D-2}} + \ldots \right] 
\; , \qquad \label{dyDelta} \\
\p^z_{\alpha} i \Delta(y;z) &\!\!\! = \!\!\!& - \frac{\Gamma(\frac{D}{2})}{2
\pi^{\frac{D}2} (a_y a_z)^{\frac{D}2 -1}} \left[  -\frac{(y \!-\! z)_{\alpha}}{
(y \!-\! z)^D} + \frac{a_z H \delta^0_{~\alpha}}{2 (y \!-\! z)^{D-2}} - 
\frac{D a_y a_z H^2 (y \!-\! z)_{\alpha}}{8 (y \!-\! z)^{D-2}} + \ldots \right]
\; . \qquad \label{dzDelta}
\end{eqnarray}
Multiplying (\ref{dyDelta}) and (\ref{dzDelta}), and retaining only potentially
divergent terms yields, 
\begin{eqnarray}
\lefteqn{\p^y_{\rho} i \Delta(y;z) \!\times\! \p^z_{\alpha} i \Delta(y;z) = 
\frac{\Gamma^2(\frac{D}{2})}{4 \pi^D (a_y a_z)^{D-2}} \Biggl\{ 
\red{-\frac{(y \!-\! z)_{\rho} (y \!-\! z)_{\alpha}}{(y \!-\! z)^{2D}}  + 
\frac{a_z H \delta_{~ \alpha}^0 (y \!-\! z)_{\rho}}{2 (y \!-\! z)^{2D-2}}} }
\nonumber \\
& & \hspace{2.5cm} \red{-\frac{a_y H \delta^0_{~\rho} (y \!-\! z)_{\alpha}}{
2 (y \!-\! z)^{2D-2}} + \frac{a_y a_z H^2 \delta_{~\rho}^0 \delta_{~\alpha}^0}{
4 (y \!-\! z)^{2D-4}}} \blue{-\frac{D a_y a_z H^2 (y \!-\! z)_{\rho} 
(y \!-\! z)_{\alpha}}{4 (y \!-\! z)^{2D-2}}} + \dots \Biggr\} \; . \qquad
\label{twoterms}
\end{eqnarray}
The terms highlighted in red and blue require separate reductions which we give 
below.

We first extract two derivatives from the red terms of expression (\ref{twoterms}),
\begin{eqnarray}
\lefteqn{\frac{\Gamma^2(\frac{D}{2})}{16 \pi^D (a_y a_z)^{D-2}} \Biggl\{ 
-\frac{(y \!-\! z)_{\rho} (y \!-\! z)_{\alpha}}{(y \!-\! z)^{2D}} + 
\frac{a_z H \delta_{~\alpha}^0 (y \!-\! z)_{\rho}}{2 (y \!-\! z)^{2D-2}}
-\frac{a_y H \delta^0_{~\rho} (y \!-\! z)_{\alpha}}{2 (y \!-\! z)^{2D-2}} 
+ \frac{a_y a_z H^2 \delta_{~\rho}^0 \delta_{~\alpha}^0}{4 (y \!-\! z)^{2D-4}}
\Biggr\} } \nonumber \\
& & \hspace{.7cm} = \frac{\Gamma^2(\frac{D}{2} \!-\! 1)}{64 \pi^D} \Biggr\{ 
\p_{\rho}^y \p_{\alpha}^z \left[ \frac{1}{(a_y a_z)^{D-2} (y \!-\! z)^{2D-4}} 
\right] + \frac{[\p_{\rho}^y \p_{\alpha}^y \!-\! \eta_{\rho\alpha} \p^2_y]}{ 
(D \!-\! 1) (a_y a_z)^{D-2}} \left[ \frac{1}{(y \!-\! z)^{2D-4}} \right] 
\Biggr\} . \qquad \label{redterms}
\end{eqnarray}
The denominator $(y - z)^{2D-4}$ is logarithmically divergent so we can extract
a local divergence from it by adding the flat space propagator equation
\cite{Onemli:2002hr,Onemli:2004mb},
\begin{equation}
\frac{1}{(y \!-\! z)^{2D-4}} = \frac{\mu^{D-4}}{2 (D \!-\! 3) (D \!-\! 4)} 
\frac{4\pi^{\frac{D}2} \, i\delta^D(y \!-\! z)}{\Gamma(\frac{D}{2} \!-\! 1)} 
- \frac{\partial^2_y}{4} \Biggl[ \frac{\ln[\mu^2 (y \!-\!z)^2]}{(y \!-\! z)^2}\Biggr]
+ O(D \!-\! 4) \; . \label{yzexpansion}
\end{equation}
Substituting (\ref{yzexpansion}) in (\ref{redterms}), and retaining only the 
divergences, reduces the red terms of expression (\ref{twoterms}) to,
\begin{eqnarray}
\lefteqn{\frac{\Gamma^2(\frac{D}{2})}{16 \pi^D (a_y a_z)^{D-2}} \Biggl\{ 
-\frac{(y \!-\! z)_{\rho} (y \!-\! z)_{\alpha}}{(y \!-\! z)^{2D}} + 
\frac{a_z H \delta_{~\alpha}^0 (y \!-\! z)_{\rho}}{2 (y \!-\! z)^{2D-2}}
-\frac{a_y H \delta^0_{~\rho} (y \!-\! z)_{\alpha}}{2 (y \!-\! z)^{2D-2}} 
+ \frac{a_y a_z H^2 \delta_{~\rho}^0 \delta_{~\alpha}^0}{4 (y \!-\! z)^{2D-4}}
\Biggr\} } \nonumber \\
& & \hspace{0cm} = \frac{\Gamma(\frac{D}{2} \!-\! 1) \mu^{D-4}}{32 \pi^{\frac{D}2}
(D \!-\! 3) (D \!-\! 4)} \Biggl\{ \p^y_{\rho} \p^z_{\alpha} \left[ 
\frac{i\delta^D(y \!-\! z)}{(a_y a_z)^{D-2}} \right] + \frac{[ \p_{\rho}^y \p_{\alpha}^y 
\!-\! \eta_{\rho\alpha} \p^2_y] \, i \delta^D(y \!-\! z)}{(D \!-\! 1) (a_y a_z)^{D-2}} 
\Biggr\} + \Bigl({\rm Finite}\Bigr) . \qquad \label{secondvanishing} 
\end{eqnarray}

We now extract derivatives from the blue term in (\ref{twoterms}),
\begin{eqnarray}
\lefteqn{-\frac{\Gamma(\frac{D}2) \Gamma(\frac{D}2 \!+\! 1) H^2}{8 \pi^D (a_y a_z)^{D-3}}
\!\times\! \frac{(y \!-\! z)_{\rho} (y \!-\! z)_{\alpha}}{(y \!-\! z)^{2D - 2}}} 
\nonumber \\
& & \hspace{2.5cm} = -\frac{\Gamma(\frac{D}{2} \!-\! 1) \Gamma(\frac{D}2 \!+\!1) H^2}{
32 \pi^D (a_y a_z)^{D-3}}  \!\times\! \left[ \frac{\p_{\rho}^y \p_{\alpha}^y}{2 (D \!-\! 3)} + 
\frac{\eta_{\rho\alpha} \p^2_y}{2 (D \!-\! 3) (D \!-\! 4)} \right] \frac{1}{(y-z)^{2D-6}} 
. \qquad \label{blueterms}
\end{eqnarray}
Of the two terms inside the square brackets at the end of (\ref{blueterms}), only the
one proportional to $\eta_{\rho\alpha} \partial^2_{y}$ is divergent. Using the same 
reduction as (\ref{yzexpansion}) we therefore reduce the blue term of expression
(\ref{twoterms}) to,
\begin{equation}
-\frac{\Gamma(\frac{D}2) \Gamma(\frac{D}2 \!+\! 1) H^2}{8 \pi^D (a_y a_z)^{D-3}}
\!\times\! \frac{(y \!-\! z)_{\rho} (y \!-\! z)_{\alpha}}{(y \!-\! z)^{2D - 2}}
= -\frac{\Gamma(\frac{D}{2} \!+\! 1)}{16 \pi^{\frac{D}2}} 
\frac{\mu^{D-4} \eta_{\rho\alpha} H^2 i \delta^D(y \!-\! z)}{(D \!-\! 3) (D \!-\! 4) 
(a_y a_z)^{D-3}} + \Bigl({\rm Finite}\Bigr) . \label{bluefinal}
\end{equation}

We are now ready to employ expressions (\ref{secondvanishing}) and (\ref{bluefinal})
in (\ref{Idef}). Note that the transverse projection operator in the second term
of (\ref{secondvanishing}) vanishes upon canceling the scale factors and exploiting
the $i \delta^D(y - z)$ to reflect derivatives where necessary, 
\begin{eqnarray}
\lefteqn{\frac{\lambda^3}{4} \partial^{\rho}_{y} \partial^{\alpha}_{z} \Biggl\{ 
(a_y a_z)^{D-2} \!\times\! \frac{\Gamma(\frac{D}2 \!-\! 1) \mu^{D-4}}{32 \pi^{\frac{D}2} 
(D \!-\! 3) (D \!-\! 4)} \frac{[\partial^{y}_{\rho} \partial^{y}_{\alpha} \!-\! 
\eta_{\rho\alpha} \partial^2_{y}] i \delta^D(y \!-\! z)}{(D \!-\! 1) (a_y a_z)^{D-2}} 
\Bigl[\delta^D(x \!-\! y) + \delta^D(x \!-\! z)\Bigr] \Biggr\} } \nonumber \\
& & \hspace{0cm} = \frac{\lambda^3 \mu^{D-4} \Gamma(\frac{D}2 \!-\! 1) 
\partial^{\rho}_{y} \partial^{\alpha}_{z}}{128 \pi^{\frac{D}2} (D\!-\!1) (D\!-\!3)
(D\!-\!4)} \Biggl\{ \Bigl[\partial^{z}_{\rho} \partial^{z}_{\alpha} \!-\! 
\eta_{\rho\alpha} \partial^2_{z} \Bigr] i \delta^D(y \!-\! z) \!\times\! 
\delta^D(x \!-\! y) \nonumber \\
& & \hspace{7cm} + \Bigl[\partial^{y}_{\rho} \partial^{y}_{\alpha} \!-\! 
\eta_{\rho\alpha} \partial^2_{y} \Bigr] i \delta^D(y \!-\! z) \!\times\! 
\delta^D(x \!-\! z) \Biggr\} = 0 \; . \qquad \label{zero}  
\end{eqnarray}
The divergent part of $I(x;y;z)$ comes from the first term of (\ref{secondvanishing}) 
and from (\ref{blueterms}). After some judicious partial integrations it can be written as,
\begin{eqnarray}
\lefteqn{ I_{\rm div} = \frac{i \lambda^3 \mu^{D-4} \Gamma(\frac{D}2 \!-\!1)}{
128 \pi^{\frac{D}2} (D \!-\! 3) (D \!-\! 4)} \Biggl\{ 2 \mathcal{D}_y \mathcal{D}_z
\left[ \frac{ \delta^D(x \!-\! y) \delta^D(x \!-\! z)}{(a_y a_z)^{D-2}} \right] -
\mathcal{D}_y \partial^{\alpha}_{z} \left[ \frac{\delta^D( y \!-\! z) \partial^z_{\alpha}
\delta^D(x \!-\! z)}{a_y^{D-2}} \right] } \nonumber \\
& & \hspace{0.5cm} - \mathcal{D}_z \partial^{\rho}_{y} \!\! \left[ \frac{\delta^D(y \!-\! z)
\partial^y_{\rho} \delta^D(x \!-\! y)}{a_z^{D-2}} \right] \!\!\Biggr\} - \frac{i \lambda^3 
H^2 a_x^2 \mu^{D-4} \Gamma(\frac{D}2 \!+\! 1)}{32 \pi^{\frac{D}2} (D \!-\! 3) (D \!-\! 4)} 
\, \partial^{\mu}_x \delta^D(x \!-\! y) \partial^x_{\mu} \delta^D(x \!-\! z) \; . \qquad 
\label{Idiv}
\end{eqnarray}

\section{Parameters and Variables }\label{variablesappendix}

The aim of this appendix is to list all the parameters and variables used in the paper. \\

In the first section, we introduce scale factor $a(t)$, Hubble parameter $H(t)$ and first slow roll parameter $\epsilon(t)$ in (\ref{geometry}). In the same section, variables $t$, $r$ and $k$ are comoving time, comoving coordiante distance to a point source and plane wave number, respectively, given in (\ref{ugrav}) and (\ref{Psigrav}). Moreover, $G$ denotes Newton's constant of gravitation. \\
In the second section, we introduce our model. $A$ and $B$ denotes two scalar fields, and the model is given in (\ref{newmodel}). $\lambda$ denotes the dimensionful coupling constant of our model. In (\ref{AandB2cterms}) and (\ref{vertcount}), the coefficients $C_{1 A^2}$, $C_{2 A^2}$, $C_{1 B^2}$, $C_{2 A^2}$, $C_{1 A B^2}$,$C_{2 A B^2}$, $C_{3 A B^2}$ and $C_{4 A B^2}$ are counterterms. Also, $D$ denotes the spacetime dimension and $\mu$ is the scale of the dimensional regularization. In the same section, we define the kinetic operator $\mathcal{D}$ in (\ref{propeq}), and provide the scalar propagator $i \Delta (x:x')$ for fields $A$ and $B$ in (\ref{propsol}). It should be noted that $k$ does not denote the wave number in the rest of the paper, but defined as a function of Hubble parameter in (\ref{propsol}). The notation adopted in the paper is summarised in the last paragraph of this section. \\
In the third section, the contribution of each diagram in Figure (\ref{fig:VEVAall}) to expectation value of $A$ is labeled $A_{2a}$ through $A_{2h}$, and the results are given in (\ref{indcont1}) and (\ref{indcont2}). Furthermore, we provide the explicit form of 3-point and 4-point contributions to 1PI 2-loop functions $-i M^2_{A}(x;x')$ and $-i M^2_{B}(x;x')$ in (\ref{A3}) through (\ref{Bc}). In the same section, $\beta$ denotes the beta function for the field $A$, and it is defined in (\ref{beta}). \\ 
Lastly in the fourth section, $\mathcal{A}$ denotes stochastic random field. In (\ref{ZAB}) and (\ref{gammas}), we provide the expressions for field strength renormalizations and gamma functions. They are denoted by $Z$ and $\gamma$, respectively. $G(x_1 ; x_2; ... ; x_n; \lambda ; \mu)$ denotes the n-point Green's function, first time appearing in (\ref{CSeqn}).


\end{document}